\documentclass[reprint, nofootinbib, amsmath,amssymb,amsfonts, aps]{revtex4-1}
\usepackage{xcolor}%Package for colours

\usepackage{bbold}
\usepackage{dsfont}
\usepackage{nccmath}
\usepackage{amsmath}
\usepackage{amssymb}
\usepackage{url}
\usepackage{colortbl}
\usepackage{multirow, tabularx}
\usepackage{threeparttable}

\newcolumntype{Y}{>{\centering\arraybackslash}X}

\def\arrvline{\hfil\kern\arraycolsep\vline\kern-\arraycolsep\hfilneg} %Vertical lines in tables
\usepackage{soul} %Strikeout

\colorlet{rowfill}{red!20}

\usepackage[titletoc,toc,title]{appendix}
\usepackage{graphicx}% Include figure files
\usepackage{dcolumn}% Align table columns on decimal point
\usepackage{bm}% bold math
\usepackage[utf8]{inputenc}
\usepackage[
]{geometry}

\usepackage{float} % Allows putting an [H] in \begin{figure} to specify the exact location of the figure
\usepackage{wrapfig} % Allows in-line images such as the example fish picture
\usepackage{caption}
\usepackage{subcaption}

\usepackage[colorlinks=true,linkcolor=red]{hyperref} %Used for hyperlinking references within own paper and  for bib
\usepackage{cleveref} %Saves time, and auto puts in correct eq, fig, e.c.t

\usepackage{float} %Override float positioning

\begin{document}
\title{Bouncing cosmologies in the presence of a Dirac-Born-Infeld field}% Force line breaks with \\

\author{Mariam Campbell$^1$, Richard Daniel$^2$, Peter Dunsby$^{1,3,4}$ and Carsten van de Bruck$^2$}

\affiliation{ }
\affiliation{$^1$Department of Mathematics and Applied Mathematics, Cosmology and Gravity Group,  University of Cape Town, Rondebosch, 7701, Cape Town, South Africa}
\affiliation{$^2$School of Mathematics and Statistics, University of Sheffield, Hounsfield Road, Sheffield S3 7RH, United Kingdom}%
\affiliation{$^3$South African Astronomical Observatory, Observatory 7925, Cape Town, South Africa}
\affiliation{$^4$Centre for Space Research, North-West University, Potchefstroom 2520, South Africa}
\vspace{5pt}

\date{\today}
\begin{abstract}
We perform a detailed dynamical system analysis for the behaviour of a Dirac-Born-Infeld (DBI) field in a spatially closed Friedmann-Lema\^itre-Robertson-Walker (FLRW) cosmology. The DBI field is characterised by a potential and brane tension. We study power-law or exponential functions for the potential and tension. 
We find that in a spatially closed FLRW cosmology, a DBI field in the ultra-relativistic limit allows for a broader range of initial conditions resulting in a bouncing universe than in the non-relativistic limit. We further note that the range of initial conditions allowing for a bounce is larger if we consider power-law functions for the potential and tension, compared to the exponential case. Our dynamical analysis shows that a DBI field does not exhibit stable cyclical behaviour, including the case in which a negative cosmological constant is present. 
\end{abstract}
\pacs{Valid PACS appear here}% PACS, the Physics and Astronomy

\maketitle

\onecolumngrid

%%%%%%%%%%%%%%%%%%%%%%%%%%%%%%%%%%%%%%%%%%%%%%%%%%%%%%%%%%%%%%%%%%%%%%%

\section{Introduction}

% Motivation for DBI
Scalar fields have been studied extensively in theories of particle physics and cosmology \cite{Kolb:1990vq, Macias:2017fpu}. In the standard model (SM), the Higgs boson is the only fundamental scalar field, but extensions of the SM, such as string theory, predict a plethora of new fundamental scalar fields \cite{Green:1987sp, Marchesano:2024gul}. In cosmology, they play an important role in models for the early and late time universe \cite{Liddle:2000cg,Amendola:2015ksp}. According to inflationary cosmology, at least one scalar field has driven a period of accelerated expansion in the very early universe and in dynamical models of dark energy the present accelerated expansion of the universe is usually driven by a scalar field. 

One interesting class of scalar fields, so--called Dirac-Born-Infeld (DBI) fields, is suggested by string theory. These fields do not possess canonical kinetic terms. They describe the motion of D--branes in a higher--dimensional warped internal space and their origin is geometrical \cite{Kachru:2003sx}. Since D--branes are described by DBI actions, it is not surprising that their kinetic term in the low--energy effective theory is non--canonical. 
The particular form of the DBI's kinetic term in the Lagrangian imposes a speed restriction analogous to the Lorentz factor in special relatively \cite{Alishahiha:2004eh,Silverstein:2003hf}.

% Successes of DBI
% Inflation
DBI fields have mainly been studied in the context of inflationary cosmology as part of $k$-inflation, inflation driven by a scalar field with a non-canonical kinetic term \cite{Armendariz-Picon:1999hyi,Garriga:1999vw,Li:2012vta}. Not surprisingly, because of the non-canonical kinetic term, the predictions of DBI inflationary models are different from standard slow-roll inflation. 
DBI's inherent speed limit allows for inflation to occur with steeper potentials compared to that of standard slow-roll inflation \cite{Chimento:2007es,Chimento:2010un,Kinney:2007ag,Bessada:2009pe,Kecskemeti:2006cg,Spalinski:2007dv}. This allows for potentials motivated by other fields of study that were ruled out otherwise, such as "stringy" potentials. However, this relativistic feature in single-field inflation amplifies non--Gaussianities compared to the standard slow--roll inflationary model. This amplification provides a possible way to constrain such models with cosmological experiments and pin down the properties and origins of an inflationary field\cite{Langlois:2008qf,Langlois:2008wt,Chen:2005fe,Weller:2011ey,Huang:2007hh,Emery:2013yua}. 

% DE
DBI fields have also been studied in the context of models for dark energy. The non-canonical kinetic term introduces interesting features in late-time cosmology, such as obtaining a cosmological constant behaviour without a false vacuum \cite{Ahn:2009xd,Ahn:2009hu,Guo:2008sz,Chimento:2009nj}. A dynamical system analysis indicates a scaling solution proven to be attractor \cite{Copeland:2010jt,Gumjudpai:2009uy,Kaeonikhom:2012xr}. However, it is found that the non-trivial solutions and critical points in the phase-space analysis are highly dependent on the Lorentz term, $\gamma$, being large. 

%Outline
In this paper we perform a detailed dynamical system analysis of DBI dark energy models and extend the work of the literature in several ways. Firstly, we allow for a non-zero spatial curvature. This is motivated by some works claiming that current observational data still marginally allows for a closed universe \cite{DiValentino:2019qzk,Yang:2022kho,Handley:2019tkm}. Even a small curvature at the present can have implications for the future evolution of the universe. Secondly, we study the impact of a negative cosmological constant on the dynamics of the DBI scalar field. There are some recent works studying dark energy in the presence of a negative cosmological constant \cite{Calderon:2020hoc,Visinelli:2019qqu,Biswas:2009fv}. From a theoretical perspective, either a non-zero curvature or a negative cosmological constant opens up the possibility of a bouncing cosmology, which could lead to a cyclic universe. Therefore, the aim of our study is to determine whether the dynamics of a DBI field naturally favours a universe with possible cyclic behaviour. 

This paper is organised as follows: in \cref{sec model}, we present the model and the autonomous set of equations for a dynamical analysis. In \cref{sec power}, we analyse the stability of a model with a power-law potential and brane tension. We also analyse the stability of an exponential behaviour in \cref{sec exp}. We illustrate the stability of both the models in \cref{sec:numerical}. We continue our analysis, examining bouncing scenarios in \cref{sec bounce} and the effect of adding a negative cosmological constant in \cref{sec neg cosmo const}. We summarise and conclude our paper in \cref{sec conc}.

% Motivation for bouncing models
% Structure

\section{The DBI model}
\label{sec model}
\noindent We consider the following action that in addition to Einstein gravity and matter, includes a DBI field, $\phi$:
\begin{equation}
    {\cal S} = \int d^4x \sqrt{-g}\left[ \frac{R}{2} - \frac{1}{f(\phi)}(\gamma^{-1}-1) - V(\phi) + {\cal L}_m \right].
\end{equation}
In this equation, ${\cal L}_m$ is the Lagrangian that encodes the matter sector, $f^{-1}(\phi)$ is the D3-brane tension, encoding geometrical properties of the bulk spacetime, and $\gamma^{-1}=\sqrt{1+fX}$ where $X=g^{\mu\nu}\partial_\mu\phi \partial_\mu\phi$ \cite{Alishahiha:2004eh,Silverstein:2003hf}. We will assume throughout this work that $V$ and $f$ are strictly positive. Adopting the Friedmann-Lema\^itre-Robertson-Walker (FLRW) metric to describe an isotropic and homogeneous universe, and in which $a(t)$ is the scale factor, $t$ the cosmic time, and assuming a homogeneous field $\phi$, the kinetic term becomes $X = -\dot\phi^2$ and therefore
\begin{equation}
    \gamma^{-1}=\sqrt{1-f\dot\phi^2},
\end{equation}
which acts as a Lorentz factor, bounding $\gamma\geq 1$ and $f^{-1}\geq\dot\phi^2$, thereby limiting the speed of the scalar field. The equation of motion for $\phi$ is (as usual $H = \dot a/a$)
\begin{equation}
    \ddot{\phi} + 3H\gamma^{-2}\dot{\phi} + \frac{1}{2}f_\phi f^{-2}(1-3\gamma^{-2} + 2\gamma^{-3}) + \gamma^{-3}V_\phi =0,
    \label{eqn metric R}
\end{equation} 
and Einstein's equations lead to the Friedmann equation:
\begin{align}
    3H^2 &= \frac{\gamma^2}{ \gamma +1 }\dot{\phi}^2 + V + \rho_m - 3\frac{K}{a^2}, \label{eqn H2}
\end{align}
where $\rho_m$ is the energy density of additional forms such as matter or radiation, and $K$ encodes the spatial curvature. We will study the case of a closed universe in this paper and set $K=1$ for the remainder of the work. 
It will also be useful to define the energy density and pressure of the DBI field:
\begin{align}
\begin{aligned}
    \rho_\phi = \frac{\gamma^2\dot{\phi}^2}{\gamma+1} + V,
    \\
    P_\phi = \frac{\gamma\dot{\phi}^2}{\gamma+1} - V,
\end{aligned}
    \label{eqn energy_pressure}
\end{align}
and $\rho_m$ and $\rho_r$ are the energy density of matter and radiation, respectively.

In the limit where $\dot\phi^2\rightarrow f^{-1}$, $\gamma^{-1}\rightarrow0$. In this limit, which we will dub as \textit{ultra-relativistic}, we see from \cref{eqn metric R} that the acceleration of field is governed by the brane tension, which results in a deceleration in an expanding universe as explored by \cite{Alishahiha:2004eh,Silverstein:2003hf}, with the Friedmann equation, \cref{eqn H2} dominated by the velocity of the field. On the other hand, when $\gamma\rightarrow1$, both \cref{eqn metric R,eqn H2} result in the equations of motion for a canonical scalar field. 

To analyse the dynamics of the system, we define the variables:
\begin{align}
    x=\frac{\dot{\phi}}{\sqrt{3\Bar{\gamma}(1+ \Bar{\gamma})}}\frac{1}{D}, ~~~y = \frac{\sqrt{V}}{\sqrt{3}D},~~~ z = \frac{H}{D},~~~ \Omega_{m} = \frac{\rho_{m}}{3D^2}
\end{align}
where 
\begin{equation}
    D = \sqrt{H^2 + \frac{K}{a^2}}, ~~~ \Bar{\gamma} = \gamma^{-1}.
\end{equation}
Here, $x$ encodes the kinetic energy of the system, $y$ the potential energy, and $\Omega_m$ is the matter density parameter, which has an equation of state parameter, $w_m$. To make the phase compact, we adopted the variable $\bar\gamma$, which is bounded between $0\leq\bar\gamma\leq1$. Furthermore, to ensure a compact phase space in a closed universe, we introduce another variable, $z$, such that the spatial curvature is encoded in the effective Hubble parameter, $D$. The Friedmann equation, Eq. \eqref{eqn H2}, in terms of these variables is then $x^{2}+y^{2}+\Omega_{m}=1$, which we will refer to as the Friedmann constraint. The energy density of the scalar field can be determined from the Friedmann constraint or Eq. $\eqref{eqn energy_pressure}$, resulting in the energy density and equation of state, 
\begin{equation}
    \Omega_\phi = x^2 + y^2, ~~ w_\phi = \frac{\bar\gamma x^2 - y^2}{x^2 + y^2},
\end{equation}
respectively. We characterise the potential and brane tension by introducing new parameters, 
\begin{equation}
    \lambda = -\frac{V_\phi}{V},~~~ \mu = -\frac{f_\phi}{f}.
\end{equation}
To ensure that the system remains closed, we introduce a new time variable, $\tau$, defined by $\frac{d}{d\tau} = D^{-1}\frac{d}{dt}$. We denote the derivative with respect to $\tau$ with a prime. The equations of motion \cref{eqn metric R,eqn H2} result in the following set of equations:
\begin{align}
    &\Bar{\gamma}' = \frac{\tilde{\gamma}\left(1-\bar{\gamma}^{2}\right)}{\sqrt{1+\tilde{\gamma}}}\left[3 \sqrt{1+\tilde{\gamma}}z+ \frac{\sqrt{3 \tilde{\gamma}}}{x}\left(\mu x^{2}-\lambda y^{2}\right)\right],\label{dynsys1}
    \\
    &x' = \frac{1}{2} \sqrt{3 \tilde{\gamma}(1+\tilde{\gamma})} \lambda y^{2} +\frac{3}{2}xz\left[-(1+\tilde{\gamma})+\left(x^{2}+y^{2}\right)\left(1+w_\phi\right) + \Omega_m(1+w_m)\right],\label{dynsys2}
    \\
    &y' = -\frac{1}{2} \sqrt{3 \tilde{\gamma}(1+\tilde{\gamma})} \lambda x y + \frac{3}{2}\left[ \left(x^{2}+y^{2}\right)\left(1+w\right) + \Omega_m(1+w_m) \right]yz,\label{dynsys3}
    \\
    &z' = (1-z^2)\left[1 - \frac{3}{2}(y^2 + x^2)(1+w) - \frac{3}{2}\Omega_m(1+w_m)\right],\label{dynsys4}
\end{align}
and for completeness,  
\begin{align}
\Omega_{m}'=-3z(1+w_{m})\Omega_{m} + 3\Omega_{m}\left[(1+w)(x^2 + y^2) + (1+w_m)\Omega_{m}\right],\label{dynsys5}
\end{align}
as a consequence of the Friedmann constraint. Maintaining generality, we also include the rate of change of $\lambda$ and $\mu$, 
\begin{align}
    \mu'&= \left(1 -\frac{f_{\phi\phi}f}{f_\phi^2} \right)\mu^2\sqrt{3\Bar{\gamma}(1+\Bar{\gamma})}x, 
    \label{eq muprime}
    \\
    \lambda'&= \left(1 -\frac{V_{\phi\phi}V}{V_\phi^2} \right)\lambda^2\sqrt{3\Bar{\gamma}(1+\Bar{\gamma})}x.
    \label{eq lambprime}
\end{align}
Utilising \cref{eq muprime,eq lambprime}, we can note that the autonomous system can describe a wide range of potentials and brane tension functions. If we are to consider an exponential model, as often explored in quintessence models, we note that the $\mu$ and $\lambda$ become a constant, reducing the number of equations to be solved.  

%%%%%%%%%%%%%%%%%%%%%%%%%%%%%%%%%%%%%%%%%%%%%%%%%%%%%%%%%%%%%%%%%%%%%%%%%%%%%%%%%%%%%%%
\section{Model with Power Law behaviour}
\label{sec power}
\noindent Following the work of \cite{Copeland:2010jt}, we consider a power law functions for both the potential $V(\phi)$ and the brane tension $f(\phi)$, which are forced to be strictly positive, 
\begin{equation}
    V(\phi) = \sigma |\phi|^p, ~~~~~~~ f(\phi) = \nu|\phi|^r. 
\end{equation}
where $(\sigma, \nu)>0$, and $p$ and $r$ are power-law parameters to be set. 
\subsection{Autonomous system}
When considering the power law functions, we re-parameterise the potential and brane tension such that,
\begin{equation}
    \Lambda = - \frac{V_\phi}{f^q V^{q+1}}, ~~~~~~~ M = -\frac{f_\phi}{f^{q+1}V^q},
    \label{eqn re-pram}
\end{equation}
where $q = -1/(p+r)$. For the power-law case we are studying here, both $\Lambda$ and $M$ are constant, which allows us to reduce the number of equations for the system. We note that the case $q =0$ identifies a specific behaviour that reduces to the exponential scenario explored later in \cref{sec exp}. 
The variables $\Lambda$ and $M$ are related to $\lambda$ and $\mu$ by  
\begin{equation}
    \lambda = \left[\frac{(1-\bar{\gamma})y^2}{\bar{\gamma}(1+\bar{\gamma})x^2}\right]^q\Lambda, ~~~~~\mu = \left[\frac{(1-\bar{\gamma})y^2}{\bar{\gamma}(1+\bar{\gamma})x^2}\right]^q M.
\end{equation}
As $\mu$ and $\lambda$ can be written in terms of $x,~y,~z$ we can utilise Eq. \eqref{eqn re-pram} with the need only to solve for $x',~y',~z'$. Equations \eqref{dynsys1}--\eqref{dynsys4} becomes:
\begin{align}
    &\bar{\gamma}' =\frac{\bar{\gamma}  \left(1-\bar{\gamma} ^2\right) }{\sqrt{\bar{\gamma} +1}} 
    \left[\frac{\bar{\gamma}^{\frac{1}{2}-q} \sqrt{3}}{x} \left(\frac{(1-\bar{\gamma} ) y^2}{ (\bar{\gamma} +1) x^2}\right)^q \left(M x^2-\Lambda  y^2\right)
    +3 \sqrt{\bar{\gamma} +1} z\right],
    \label{eq gamma_q}
    \\
    &x' = \frac{\sqrt{3}}{2}\left[\bar{\gamma}(\bar{\gamma} +1)\right]^{\frac{1}{2}-q}  \left[\frac{(1-\bar{\gamma}) y^2}{   x^2}\right]^q  \Lambda  y^2+\frac{3xz}{2}\left[ (\bar{\gamma} +1)( x^2 -1) +\Omega_m(1+w_m)\right],
    \label{eq x_q}
    \\
    &y' = -\frac{\sqrt{3}}{2}\left[\bar{\gamma} (\bar{\gamma} +1)\right]^{\frac{1}{2}-q} \left[\frac{(1-\bar{\gamma}) y^2}{  x^2}\right]^q \Lambda xy + \frac{3}{2}yz \left[x^2(1+\bar{\gamma}) + \Omega_m(1+w_m)\right],
    \label{eq y_q}
    \\
    &z' = \left(1-z^2\right) \left[1-\frac{3}{2} x^2 \left(\bar{\gamma}+1\right) - \frac{3}{2}\Omega_m(1+w_m) \right].
    \label{eq z_q}
\end{align}

\subsection{Stability analysis}
The equations \eqref{eq gamma_q} to \eqref{eq y_q} potentially contains a singular fixed point due to $\bar\gamma^{-1}$ and $x^{-1}$ terms. Therefore, in order to examine the fixed points, we determine the fixed points for $\bar\gamma$,
\begin{align*}
    &\bar{\gamma} = 0, 
    \\
    &\bar{\gamma} =1,
    \\ 
    &\frac{\bar{\gamma}^{\frac{1}{2}-q} \sqrt{3}}{x} \left(\frac{(1-\bar{\gamma} ) y^2}{  (\bar{\gamma} +1) x^2}\right)^q \left(M x^2-\Lambda  y^2\right)
    +3 \sqrt{\bar{\gamma} +1} z =0.
\end{align*}
We also restrict ourselves to examining unambiguous fixed points (solutions that do not result in 0/0) at the background level. 
We then examine the range of fixed points for $x,y,z$ for each case, summarised in \cref{tab:power}. 

\subsubsection*{Case I: $\bar\gamma=0$}
The fixed point $\Bar{\gamma}=0$ corresponds to an ultra-relativistic field, with $\gamma\rightarrow\infty$. Therefore, the fixed point is an asymptotic solution instead of a truly physical one. In the ultra-relativistic scenario, we find a condition on $q$, $q\leq1/2$, such that Eqs. \eqref{eq gamma_q}--\eqref{eq y_q} remains physical, thus constraining the potentials and warp functions. The resulting set of equations becomes
\begin{align}
    &x' = \frac{3xz}{2}\left[( x^2 -1) +\Omega_m(1+w_m)\right],
    \label{eq q0 x}
    \\
    &y' = \frac{3}{2}yz \left[x^2 + \Omega_m(1+w_m)\right],
    \label{eq q0 y}
    \\
    &z' = \left(1-z^2\right) \left[1-\frac{3}{2} x^2 - \frac{3}{2}\Omega_m(1+w_m) \right].
    \label{eq q0 z}
\end{align}
In this case, the equations are now void of the parameter $q$, where the condition $q\leq1/2$ has to be satisfied so that the system remains physical. We also note, given this condition, the equations are independent of the variable $M$ and $\Lambda$, and thus on the specific choice of $V$ and $f$.

Fig. \ref{fig:power-law stability g0}, panels P1--P5, describes the stability of the ultra-relativistic case for fixed points indicating expansion, contraction and at the bounce.
Panels P1 and P3 shows the stability of the fixed points describing contraction, i.e. $z=-1$. When considering ultra-stiff matter, $w_{m}=1$, and $\Omega_{m}>0.5$ we find the stability of these fixed points to be a saddle. When $w_{m}=1/3$, radiation filled, and $\Omega<0.7$, we see that the stability for P1 is stable and P3 is unstable.\\
Panels P2 and P4 shows the stability of the fixed points describing expansion, i.e. $z=1$. During expansion, we see that the stability of P2 is a swap of the stability of P1 and the stability of P4 is a swap of the stability of P3 for the matter cases considered.\\
Panel P5 describes the stability at the bounce, i.e. $z=0$. We see that when $w_{m}=1/3$, radiation filled, with $\Omega_{m}<0.5$ the bounce has a saddle stability. Pressureless matter, $w_{m}=0$, with $\Omega<0.65$, shares the same stability. When considering ultra-stiff matter, $w_{m}=1$, and $\Omega_{m}>0.3$, no real fixed points exist.

\subsubsection*{Case II: $\bar\gamma=1$}
In this scenario, provided $q\neq0$, the dependence on $\gamma$ is removed. In the case where $q=0$, the system of equations reduces back to Eqs. \eqref{dynsys1}--\eqref{dynsys4}. The $q=0$ scenario, is further explored in \cref{sec exp} with the resulting fixed points given in \cref{expo-potential-stability}.
Many fixed points will result in an ambiguity, i.e terms that contain $0/0$, such as $x\rightarrow0$. These are treated with care, and the non-divergent solutions are listed in \cref{tab:power}. The autonomous system is, for $q\neq0$,
\begin{align}
    &x' = \frac{3xz}{2}\left[ 2( x^2 -1) +\Omega_m(1+w_m)\right],
    \label{eq x_q1}
    \\
    &y' = \frac{3}{2}yz \left[2x^2 + \Omega_m(1+w_m)\right],
    \label{eq y_q1}
    \\
    &z' = \left(1-z^2\right) \left[1-3 x^2  - \frac{3}{2}\Omega_m(1+w_m) \right].
\end{align}

It is worth noting that for $\bar\gamma=1$ the DBI field behaves identically to the usual canonical scalar field. Interpreting $z>0$ as a strictly expanding universe, the fixed points are $x=1,y=0, w_m=-1$, corresponding to the power-law inflationary fixed points \cite{Lucchin:1984yf}. Therefore, in later sections, we will compare the effects of including $\bar\gamma\neq 1$ against a canonical scalar field, $\bar\gamma= 1$.

We find that the ultra-relativistic and the canonical scalar have very similar resulting equations of motion, and as such similar behaving fixed points. For instance, in regards to $z'$, the fixed point is determined by the kinetic energy, the value of $x^2$ as seen from \cref{eq z_q}. From \cref{tab:power}, we see that when the relative kinetic energy of our scalar field is not dominating, $x^2\rightarrow 0$, we find the dynamics result in a quasi-de-sitter expansion. An expected result from a slow-rolling scalar field. However, with a fast-rolling scalar field, $x^2\rightarrow 1$, the dynamics result in a singular collapse, identified as the "Big Crunch".

\begin{figure}[h]
    \centering
    \includegraphics[scale=0.85]{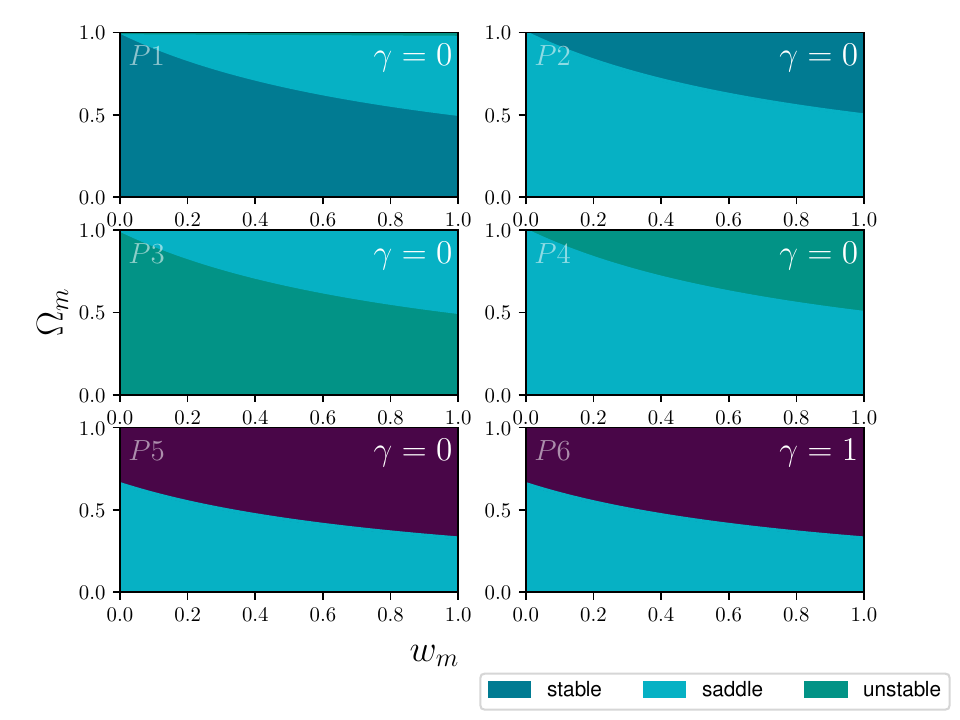}
    \caption{Stability of fixed points for the power-law potential case, with $q=\frac{1}{2}$, as referenced in Table \ref{tab:power}. The purple regions in $P5$ and $P6$ correspond to a pair of parameter values that result in an imaginary fixed point.}
    \label{fig:power-law stability g0}
\end{figure}

Fig. \ref{fig:power-law stability g0} P6, shows the stability at the bounce, $z=0$, for the specific case $q=1/2$. The stability at the bounce is the same as in the ultra-relativistic case, $\bar{\gamma}=0$, at the bounce.

\subsubsection*{Case III: $0<\gamma<1$}
\noindent Here we further identify fixed points using $\frac{\bar{\gamma}^{\frac{1}{2}-q} \sqrt{3}}{x} \left(\frac{(1-\bar{\gamma} ) y^2}{  (\bar{\gamma} +1) x^2}\right)^q \left(M x^2-\Lambda  y^2\right)
    +3 \sqrt{\bar{\gamma} +1} z =0$ where $\bar{\gamma}$ remains a constant between 0 and 1. Assuming that $q\neq0$, we use the fixed point to remove the dependence on the $q$--parameter by setting
\begin{equation}
    \sqrt{3}\left[\bar\gamma(\bar\gamma +1) \right]^{\frac{1}{2}-q}\left[\frac{(1-\bar{\gamma}) y^2}{   x^2}\right]^q = \frac{3xz \sqrt{\bar\gamma +1}}{(\Lambda y^2 - Mx^2)}.
\end{equation}
This allows us to remove the $q$ dependency in Eqs. \eqref{eq x_q} and \eqref{eq y_q}:
\begin{align}
    &x' = \frac{3}{2}\left[\frac{\sqrt{\bar\gamma +1}}{(\Lambda y^2 - Mx^2)}  \Lambda  y^2+ (\bar{\gamma} +1)( x^2 -1) +\Omega_m(1+w_m) \right]xz,
    \label{eqn gx}
    \\
    &y' = \frac{3}{2}\left[-\frac{ \sqrt{\bar\gamma +1}}{(\Lambda y^2 - Mx^2)} \Lambda x^2  + (1+\bar{\gamma})x^2 +\Omega_m(1+w_m)\right]yz,
    \label{eqn gy}
\end{align}
with $z'$ remaining unchanged. Upon analysing the set of equations \eqref{eq z_q}, \eqref{eqn gx} and \eqref{eqn gy}, we find that there are no stable fixed points for $0<\bar\gamma<1$, as identified in \cref{tab:power}. Regarding the resulting behaviour of the system, we utilise the relationship between $z'$ and $x^2$. If the kinetic energy is dominating in \cref{eq z_q} such that $x^2 \left(\bar{\gamma}+1\right) > \frac{2}{3}- \Omega_m(1+w_m) $ this drives $z$ to negative values. On the other hand, if the kinetic energy is not dominating, $x^2 \left(\bar{\gamma}+1\right) < \frac{2}{3}- \Omega_m(1+w_m) $, then $z$ will be driven to positive values. 

Furthermore, if we allow $\bar\gamma$ to evolve from \cref{dynsys1}, we can see that the resulting negative (positive) $z$, which is a product of large (small) values of $x^2$ drives $\bar\gamma$ towards zero (one) and consequently to the fixed points determined for $\bar\gamma =0$ or $\bar\gamma=1$. We interpret this result as the system being highly unstable in the regime of $0<\bar\gamma<1$, with the resulting stable fixed point highly dependent on the initial conditions given.

\begin{table}[!htb]
  \centering
  \hspace{-1cm}
  \begin{tabular}{|c|c|c|c|l|c|c|c|c|}
    \hline
    {$\gamma$} & {$x$} & {$y$} & {$z$} &
    \multicolumn{2}{c|}{Existence} & {Stability} \\
    \hline
    0 & $\pm\sqrt{\frac{2 - 3\Omega_m(1+w_m)}{3}}$ & $\sqrt{\frac{1}{3} + \Omega_mw_m}$ & 0 & \multicolumn{2}{l|}{$q\leq 1/2$  $\Omega_m(1+w_m)\leq3/2$} & saddle\\
    \cline{5-7}
    &&&& $\Omega_{m}\in(0:1)$ & $w_{m}\in(0:1)$ & Fig. \ref{fig:power-law stability g0}, P5\\
    \hline
    0 & $\pm\sqrt{1-\Omega_m(1+w_m)}$ & 0 & -1 & \multicolumn{2}{l|}{$q\leq 1/2, ~\Omega_mw_m=0$ , $\Omega_m(1+w_m)\leq 1$} & stable\\
    \cline{5-7}
    &&&& $\Omega_{m}\in(0:1)$ & $w_{m}\in(0:1)$ & Fig. \ref{fig:power-law stability g0}, P1\\
    \hline
    0 & $\pm\sqrt{1-\Omega_m(1+w_m)}$ & 0 & 1 & \multicolumn{2}{l|}{$q\leq 1/2, ~\Omega_mw_m=0$ , $\Omega_m(1+w_m)\leq 1$} & unstable\\
    \cline{5-7}
    &&&& $\Omega_{m}\in(0:1)$ & $w_{m}\in(0:1)$ & Fig. \ref{fig:power-law stability g0}, P2\\
    \hline
    0 & 0 & $\sqrt{1-\Omega_m(1+w_m)}$ & -1 & \multicolumn{2}{l|}{$q\leq 1/2, ~\Omega_mw_m=0$ , $\Omega_m(1+w_m)\leq 1$} & unstable\\
    \hline
    0 & 0 & $\sqrt{1-\Omega_m(1+w_m)}$ & 1 & \multicolumn{2}{l|}{$q\leq 1/2, ~\Omega_mw_m=0$ , $\Omega_m(1+w_m)\leq 1$} & stable\\
    \hline
    0 & 0 & 0 & -1 & \multirow{2}{*}{$q\leq1/2$, $\Omega_{m}=1$} & $w_{m}>0$ & stable\\
    \cline{6-7}
    &&&&& $w_{m}<0$ & unstable\\
    \cline{5-7}
    &&&& $\Omega_{m}\in(0:1)$ & $w_{m}\in(0:1)$ & Fig. \ref{fig:power-law stability g0}, P3\\
    \hline
    0 & 0 & 0 & 1 & \multirow{2}{*}{$q\leq1/2$, $\Omega_{m}=1$} & $w_{m}>0$ & unstable\\
    \cline{6-7}
    &&&&& $w_{m}<0$ & stable\\
    \cline{5-7}
    &&&& $\Omega_{m}\in(0:1)$ & $w_{m}\in(0:1)$ & Fig. \ref{fig:power-law stability g0}, P4\\
    \hline
    1 & $\pm\sqrt{\frac{2 - 3\Omega_m(1+w_m)}{6}}$ & $ \sqrt{\frac{4-3\Omega_m(1-w_m)}{6}}$ & 0 & \multicolumn{2}{l|}{$q>0 \in \mathbb{R}$ , $\Omega_m(1+w_m)\leq2/3$} & saddle\\
    \cline{5-7}
    &&&& $\Omega_{m}\in(0:1)$ & $w_{m}\in(0:1)$ & Fig. \ref{fig:power-law stability g0}, P6\\
    \hline
    1 & $\pm\frac{\sqrt{2-\Omega_m(1+w_m)}}{\sqrt{2}}$ & 0 & -1 & \multicolumn{2}{l|}{$q>0,~ \Omega_m(1-w_m)=0,~ \Omega_m(1+w_m)\leq 2$} & stable\\
    \hline
    1 & $\pm\frac{\sqrt{2-\Omega_m(1+w_m)}}{\sqrt{2}}$ & 0 & 1 & \multicolumn{2}{l|}{$q>0,~ \Omega_m(1-w_m)=0,~ \Omega_m(1+w_m)\leq 2$} & unstable\\
    \hline
    1 & 0 & $\sqrt{1-\Omega_m}$ & 1 & \multicolumn{2}{l|}{$ q>0,~ w_m=-1$} & stable\\
    \hline
    1 & 0 & $\sqrt{1-\Omega_m}$ & -1 & \multicolumn{2}{l|}{$ q>0,~ w_m=-1$} & unstable\\
    \hline
    1 & 0 & 0 & -1 & \multirow{3}{*}{$q>0$, $\Omega_{m}=1$} & $w_{m}=1$ & stable\\
    \cline{6-7}
    &&&&& $|w_{m}|<1$ & saddle\\
    \cline{6-7}
    &&&&& $w_{m}=-1$ & unstable\\
    \hline
    1 & 0 & 0 & 1 & \multirow{3}{*}{$q>0$, $\Omega_{m}=1$} & $w_{m}=1$ & stable\\
    \cline{6-7}
    &&&&& $|w_{m}|<1$ & saddle\\
    \cline{6-7}
    &&&&& $w_{m}=-1$ & unstable\\
    \hline
    $\frac{1+3w_m\Omega_m}{3(\Omega_m-1)}$ & $\pm\sqrt{1-\Omega_m}$ & 0 & 0 & \multicolumn{2}{l|}{$q>0, ~w_m<0,~\Omega_m>\frac{1}{3}$, $\frac{2}{3}>\Omega_m(1+w_m)$} & unstable\\
    \hline
    $\frac{3-\Lambda^2(1-\Omega)}{3}$ & 0 & $\sqrt{1-\Omega_m}$ & $\pm1$ & \multicolumn{2}{l|}{$q=-\frac{1}{2},~\Lambda^2(1-\Omega_m)<3$} & unstable\\
    \hline
  \end{tabular}
  \caption{\label{tab:power}Each of the physical fixed points with corresponding stability and conditions for existence, for the power-law potential case.}
\end{table}

%%%%%%%%%%%%%%%%%%%%%%%%%%%%%%%%%%%%%%%%%%%%%%%%%%%%%%%%%%%%%%%%%%%%%%%%%%%%%%%%%%%%%%%

\section{Model with an exponential potential}
\label{sec exp}
\noindent 
Now we consider the case in which both the potential and brane tension depend exponentially on the scalar field, 
\begin{equation}
    V(\phi)=\sigma e^{-\lambda\phi}, \hspace{1cm} f(\phi)=\nu e^{-\mu\phi}, 
\end{equation}
where $\sigma$, $\nu$, $\mu$ and $\lambda$ are constants. This implies that \cref{eq muprime,eq lambprime} are automatically set to zero. The power-law case in \cref{sec power} can be related to an exponential function with $\phi$ in the limit $(p,r)\rightarrow\infty$. This reduces the following parameters to $q\rightarrow0$ and $\Lambda \rightarrow\lambda$, $M\rightarrow\mu$. The resulting autonomous equations are given by Eqs. \eqref{dynsys1}--\eqref{dynsys4}. In the limit $q=0$, the autonomous system is described by Eqs. \eqref{eq gamma_q}--\eqref{eq z_q}.

\subsection{Stability analysis}
Performing the same fixed point analysis as in \cref{sec power}, we begin by determining the fixed points of Eq. \eqref{dynsys1}, which has three roots:
\begin{align*}
    &\Bar{\gamma}=0,\\
    &\Bar{\gamma}=1,\\
    &\Bar{\gamma}=\frac{3 x^2 z^2}{(\mu x^2 - \lambda y^2)^2-3 x^2 z^2 }.
\end{align*}
\subsubsection*{Case I: $\bar\gamma=0$}
\noindent Setting $q=0$ for $\bar\gamma =0$ results in the same set of autonomous equations as Eqs. \eqref{eq q0 x}--\eqref{eq q0 z}. Therefore, the fixed points and their stability are the same in this case as given in \cref{tab:power}. 
\subsubsection*{Case II: $\bar\gamma=1$}
\noindent Re-emphasising, as in \cref{sec power}, $\Bar{\gamma}=1$ represents a canonical scalar field. For the case $q=0$, the result is an exponential potential similar to that found in dark energy models \cite{Copeland:1997et}. The autonomous system reads, 
\begin{align}
    &x' = \frac{3}{2}xz\left[2(x^2-1) + \Omega_m(1+w_m)\right] + \sqrt{\frac{3}{2}} \lambda y^{2},
    \\
    &y' = \frac{3yz}{2}\left[2x^2 + \Omega_m(1+w_m) \right] - \sqrt{\frac{3}{2}} \lambda x y,
    \\
    &z' = (1-z^2)\left[1 - 3x^2 - \frac{3}{2}\Omega_m(1+w_m)\right].
\end{align}
The fixed points' outcomes and characteristics are presented in \cref{expo-potential-stability}. The table shows that the fixed points exhibit behaviour analogous to the power law's description. Specifically, when the potential is zero ($y=0$), the only stable fixed point corresponds to a collapse leading to a singular "Big Crunch." However, including a non-zero potential ($y>0$) results in the general behaviour of the Universe, which tends toward a quasi-de-Sitter solution. This finding aligns with prior research in the field \cite{Copeland:1997et,Paliathanasis:2015gga,Copeland:2006wr}. It tells us that the universe will eventually collapse without a scalar field i.e. no dark energy or a zero potential. However, it is important to acknowledge that these scenarios are physically unrealistic, as dictated by the inherent nature of exponential decay where $V>0$, implying $y\neq0$. The solution resembles standard quintessence with a non-zero potential, with dark energy dominating the dynamics.

\subsubsection*{Case III: $\Bar{\gamma}=\frac{3 x^2 z^2}{-3 x^2 z^2 + (\mu x^2 -\lambda y^2)^2}$}
The resulting fixed points specifying $\Bar{\gamma}=\frac{3 x^2 z^2}{-3 x^2 z^2 + (\mu x^2 -\lambda y^2)^2}$ result in either $x=0$ or $z=0$. Therefore, $\Bar{\gamma} = 0$, resulting in a repeated fixed point. Thus, the $q=0$ case results in no new fixed points.

%%==== table begins for exponential potential ====%
\begin{table}
\centering
\begin{tabularx}{\textwidth}{|c|c|c|c|c|X|c|}
\hline
$~\Bar{\gamma}~$ & $x$ & $y$ & $z$ & $\Omega_{m}$ & Existence & Stability \\ \hline \hline
1 & $-1$ & 0 & $-1$ & 0 & $0\leq w_{m}<1$ and $0\leq\lambda\leq\frac{6}{\sqrt{6}}$ & stable \\ 
\cline{6-7}
& & & & & $\hspace{1em}0\leq w_{m}\leq1$ and $\lambda>\frac{6}{\sqrt{6}}$ & saddle \\
\hline
$1$ & $-1$ & $0$ & $1$ & $0$ & $0\leq w_{m}\leq1$ and $\lambda>0$ & unstable\\
\hline
$1$ & $0$ & $0$ & $-1$ & $1$ & $0\leq w_{m}<1$ and $\lambda\in\mathbb{R}$ & saddle\\
\cline{6-7} &&&&& $w_{m}=1$ and $\lambda\in\mathbb{R}$ & stable\\
\hline
$1$ & $0$ & $0$ & $1$ & $1$ & $0\leq w_{m}<1$ and $\lambda\in\mathbb{R}$ & saddle\\
\cline{6-7} &&&&& $w_{m}=1$ and $\lambda\in\mathbb{R}$ & unstable\\
\hline
$1$ & $1$ & $0$ & $-1$ & $0$ & $0\leq w_{m}\leq1$ and $\lambda\in\mathbb{R}$ & stable\\
\hline
$1$ & $1$ & $0$ & $1$ & $0$ & $0\leq w_{m}\leq1$ and $0\leq\lambda<\frac{6}{\sqrt{6}}$ & unstable\\
\hline
$1$ & $-\frac{1}{\sqrt{3}}$ & $\sqrt{\frac{2}{3}}$ & $-\frac{\lambda}{\sqrt{2}}$ & $0$ & $0\leq w_{m}\leq1$ and $\sqrt{\frac{6}{3}}<\lambda\leq\sqrt{\frac{8}{3}}$ & unstable\\
\cline{6-7} &&&&& $0\leq w_{m}\leq1$ and $0<\lambda<\sqrt{2}$ & saddle\\
\cline{6-7} &&&&& $0\leq w_{m}\leq1$ and $\lambda>\sqrt{\frac{8}{3}}$ & unstable spiral\\
\hline
$1$ & $\frac{1}{\sqrt{3}}$ & $\sqrt{\frac{2}{3}}$ & $\frac{\lambda}{\sqrt{2}}$ & $0$ & $0\leq w_{m}\leq1$ and $\sqrt{2}<\lambda\leq\sqrt{\frac{8}{3}}$ & stable\\
\cline{6-7} &&&&& $0\leq w_{m}\leq1$ and $0<\lambda<\sqrt{2}$ & saddle\\
\cline{6-7} &&&&& $0\leq w_{m}\leq1$ and $\lambda>\sqrt{\frac{8}{3}}$ & unstable spiral\\
\hline
\end{tabularx}
\caption{Additional fixed points for $q=0$, an exponential potential and brane tension.}
\label{expo-potential-stability}
\end{table}

%_____________________________________________________________________________________________________________________________________________________________________________________

\section{Numerical analysis}\label{sec:numerical}
\begin{figure}[ht!]
    % \centering
    % \begin{subfigure}[b]{0.49\textwidth}
    %     \includegraphics[width=\textwidth]{dyn_q=0.0.png}
    %     \label{fig dyn_q=0}
    %     \caption{$q=0$}
    % \end{subfigure}
    % %
        \includegraphics[width=\textwidth]{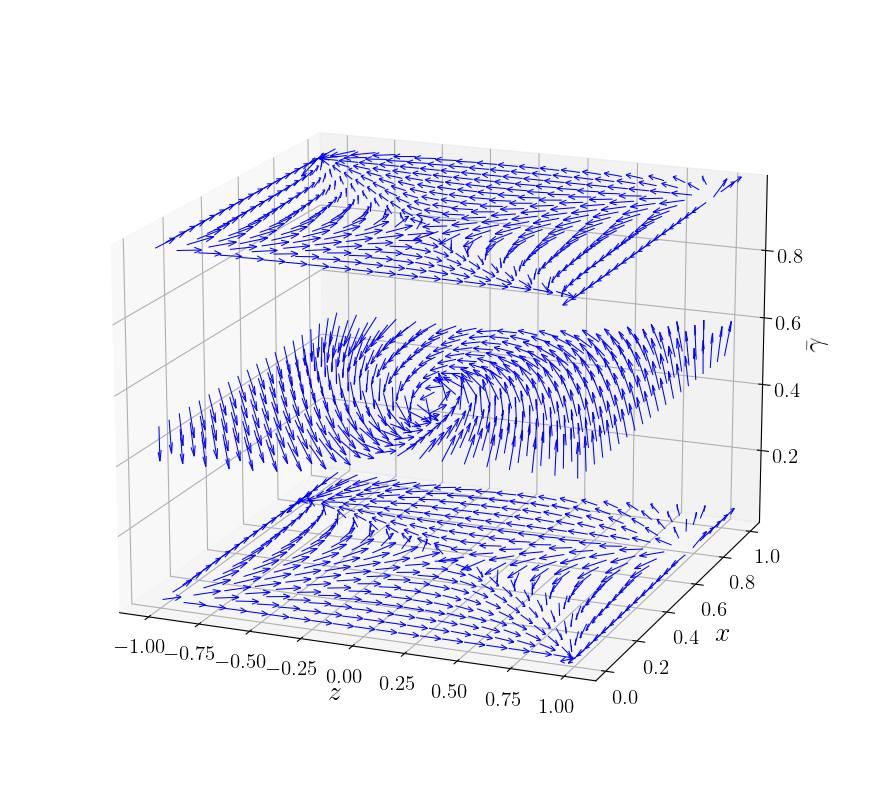}
    \caption{A phase plot showing the behaviour of the system with $q=1/2$, $\Omega_m = 0.2$, $w_m=1/3$ and $M=\Lambda=10^{-4}$.}
    \label{fig dynamical_sys}
\end{figure}
\begin{figure}[ht!]
        \includegraphics[width=\textwidth]{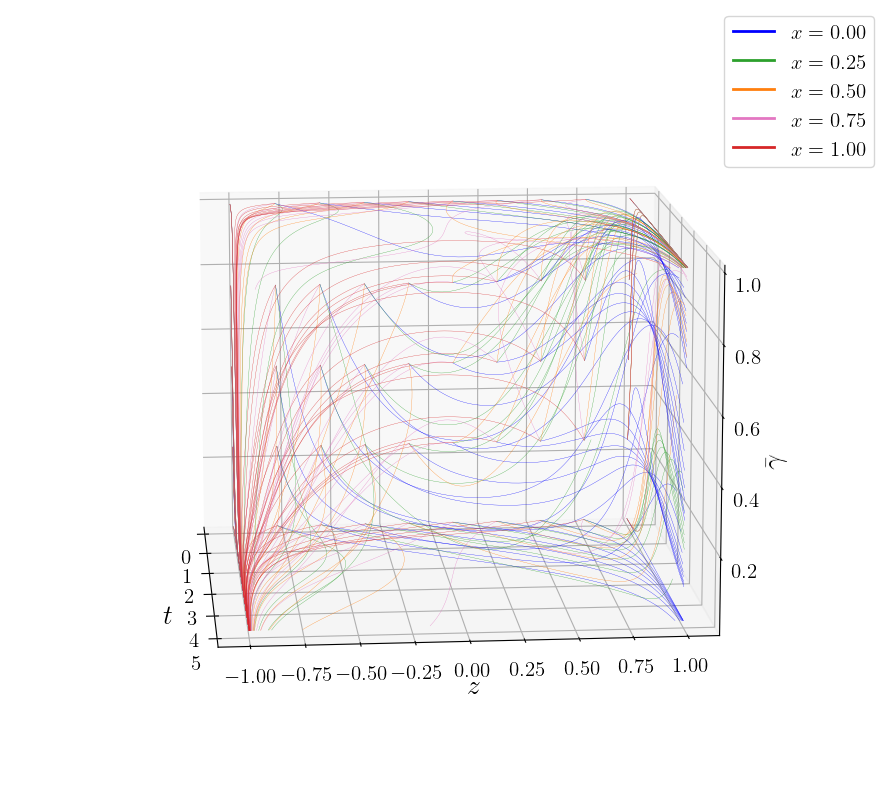}
    \caption{Phase plots analysing the role of initial conditions for the power law model with $q=1/2$. The other model parameter are set to $M=\Lambda = 10^{-4}$ and $\Omega_m=0$. The colours of the trajectories indicate the initial value of $x$, as shown in the legend. It can be seen that trajectories with larger initial values of $x$ end up at $z=-1$, 
    whereas trajectories with smaller initial values of $x$ end up at $z=1$.}
    \label{fig q_change}
\end{figure}

\noindent We numerically analyse the system's behaviour for the case $q=1/2$ to indicate the behaviour of the commonly cited scenario of $V\propto \phi^2$ and $f\propto \phi^{-4}$. We have numerically explored other values of $q$, with little difference in the overall dynamical behaviour. We also note that the behaviour of the potential, $y$, is not included in \cref{fig dynamical_sys,fig q_change} as this provides little insight and is constrained by the Friedmann equation. 

The fixed points resulting in $\bar\gamma=0$ and $\bar\gamma =1$ have similar behaviours, as illustrated in Fig. \ref{fig dynamical_sys}. We see the emergence of the stable solutions, identified in \cref{tab:power}, for ${y, z}={0, -1}$ and ${x, z}={0, 1}$, which correspond to an accelerated collapse or accelerated expansion, respectively. The saddle solution at $z=0$ is also clearly depicted in Fig. \ref{fig dynamical_sys}, for $\bar\gamma=0$ and $\bar\gamma =1$, corresponding to the saddle found in \cref{tab:power}.

Although, there seems to be a resemblance of an almost unstable cyclic behaviour at $z=0, ~\Bar{\gamma}=1/2$, this is an artefact of the transition between  $z<0$ flowing towards $\bar\gamma\rightarrow 0$ and $z>0$ flowing towards $\bar\gamma\rightarrow 1$. There is an unstable point at $z=0$ while $0<\bar{\gamma}<1$ in \cref{tab:power}, but  it is not well-defined given the choice of parameters in Fig. \ref{fig dynamical_sys}. Expanding the cross-section to examine further values of $\bar\gamma$ between $0$ and $1$, result in the same behaviour as seen for $\bar\gamma = 1/2$. We conclude that the dynamics in the regime $1>\bar\gamma>0$ quickly result in an ultra-relativistic $\Bar{\gamma}=0$ or standard quintessence $\Bar{\gamma}=1$ scenario. Thus, the dynamics always result in a quasi-de-Sitter Universe, $z\rightarrow\pm 1$, as specified by stable solutions in \cref{tab:power}.

We also study the initial conditions and the time taken for the Universe to end up in the final solution of $z\rightarrow\pm1$ as shown in Fig. \ref{fig q_change}, where the colours of the individual trajectories denote the initial value for the variable $x$. As it can be seen, trajectories in phase space are highly dependent on the initial conditions, which is similar to what happens in many models that include bouncing cosmologies. A universe is generally more likely to end up in an expanding Universe if it has a dominant kinetic energy and more likely to result in a collapse if the kinetic energy is subdominant. This effect can be seen in Fig. \ref{fig q_change}, where the initial conditions of the kinetic term vary in values of $\bar{\gamma}$. In addition, there are unstable spiral solutions that exhibit a cyclic solution which quickly tend towards and singular collapse or quasi-de-Sitter Universe. However, these are not included in Fig. \ref{fig q_change}, due to the resolution and the exhaustive numerical analysis required to study these particular solutions. 

An interesting result from our analysis is the enhancement of the phase space of a resulting bounce scenario ($z<0 \rightarrow z>0$) in ultra-relativistic scenarios compared to a canonical scalar field scenario. This feature is explored further in the next section.

%%%%%%%%%%%%%%%%%%%%%%%%%%%%%%%%%%%%%%%%%%%%%%%%%%%%%%%%%%%%%%%%

\section{Bouncing cosmologies with a DBI field}
\label{sec bounce}
We have concluded that stable, generic cyclic models are not exhibited in a DBI model. However, it was found that the "deceleration" mechanisms of the scalar field provide a more extensive range of initial conditions that lead to a bounce. This can be seen in figures \cref{fig dynamical_sys} and \cref{fig dyn_sys_lamb} for the ultra-relativistic case ($\bar\gamma=0$), compared to the standard quintessence case ($\bar\gamma =1$). 

\begin{figure}[ht]
    \centering
        \includegraphics[width=\textwidth]{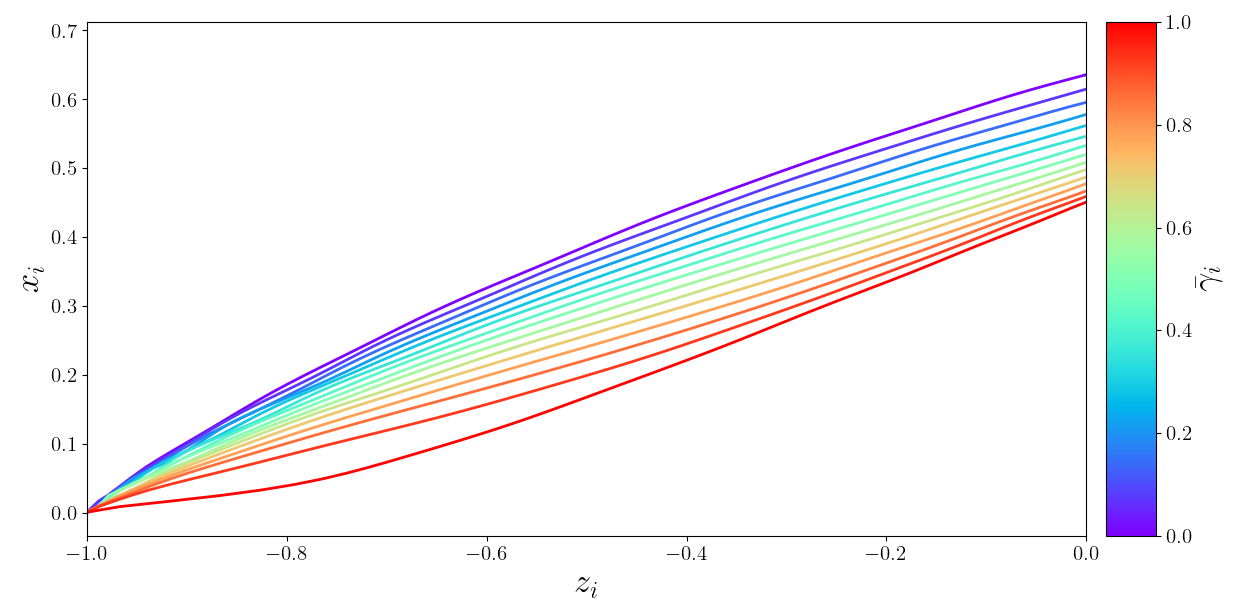}
    \caption{Lines identify the maximum initial condition that results in an expanding universe, $z>0$, with each colour identified as the initial value of $\bar\gamma$, $\bar\gamma_i$. Therefore, the area under the line for each case of $\bar\gamma_i$ encodes the initial conditions that result in a bouncing scenario.}
    \label{fig dyn_IC}
\end{figure}

We investigate this aspect further in the following. Provided a range of initial conditions for $z_i$, $x_i$, with the dependence on $y_{i}$ determined by the Friedmann constraint, we numerically integrate the system as before for \cref{fig q_change}. This results in a relationship between the initial conditions and the dynamics, resulting in a bounce. This is shown in \cref{fig dyn_IC}, in which the area under each graph shows the initial conditions that will result in a bounce.

In the case of an exponential form for potential and brane tension, for which the results are shown in Fig. \ref{fig dyn_IC}, we find an increase of $69\%$ in initial conditions for the ultra-relativistic case, $\bar\gamma=0$, will lead to a bounce, whereas $\bar\gamma=0.5$ leads to a $36\%$ increase in initial conditions resulting in a bounce. For the power-law case with $q=0.5$, we find that for the ultra-relativistic case, there is a $77\%$ increase in the initial conditions that result in a bounce. This drops to a $41\%$ increase at $\bar\gamma=0.5$.  

Note that smaller $z_i$ values correspond to a slower contraction of the Universe, which results in more time for the spatial curvature to dominate and a larger range of $x_i$ values that result in a bounce. We can see from Fig. \ref{fig dyn_IC}, that including the DBI model with the condition $\bar{\gamma}<1$, the range of initial conditions that lead to a bounce increases as the initial value of $\Bar{\gamma}$ grows towards an ultra-relativistic state, $\bar\gamma=0$.

%%%%%%%%%%%%%%%%%%%%%%%%%%%%%%%%%%%%%%%%%%%%%%%%%%%%%%%%%%%%%%%%

\newpage
\section{Adding a negative cosmological constant}
\label{sec neg cosmo const}
As we have seen in the last section, after the bounce the dynamics results in a quasi-de-Sitter expansion, $z\rightarrow +1$. In the following we include a negative cosmological constant to arrange for this system to re-collapse. The aim of this analysis is to see whether universe would as a result, settle into a cyclic behaviour.

We include a cosmological constant denoted by an energy density $\rho_C$ (the equation of state is -1). The Friedmann equation becomes,
\begin{equation}
    3H^{2}=\frac{\gamma^{2}\dot{\phi}^2}{\gamma+1}+V-3\frac{K}{a^{2}}-\rho_{C}.
\end{equation}
In the following we define the quantity $\tilde{D}^{2}=H^{2}+\frac{K}{a^{2}}+\frac{\rho_{C}}{3}$ in order to compactify the $K>0$ state space. In addition, we define a new dimensionless variable $\Omega_{C}\equiv\frac{\rho_{C}}{3\tilde{D}^{2}}$. The only change to the equations of motion Eqs. \eqref{eq gamma_q}--\eqref{eq z_q} is 

\begin{align}
    {z}' = (1-{z}^2)\left[1 - \frac{3}{2}x^2(1+\bar{\gamma})\right] - \Omega_{C}\label{dynsys13},
\end{align}

where $^\prime \equiv \tilde{D}^{-1}\frac{d}{dt}$ as before, and the Friedmann contraint remains $x^{2}+y^{2} + \Omega_m=1$. The same analysis can be performed by setting $w_m=-1$, which will remove $\Omega_m$ from Eqs. \eqref{eq gamma_q}--\eqref{eq z_q}, giving the same results. We choose the equations above, as it provides an explicit dependence on $\Omega_C$. We furthermore consider the power law case with $q=1/2$.
\begin{table}
\hspace*{-1cm}
\centering
\begin{tabularx}{1.15\textwidth}{|c|c|c|c|X|c|}
\hline
$~~\Bar{\gamma}~~$ & $x$ & $y$ & $z$ & ~~Existence & ~~Stability~~ \\ \hline \hline
0 & 0 & $\pm\sqrt{1-\Omega_m}$ & $-\sqrt{\frac{2(\Omega_C -1)+ 3\Omega_m(1+w)}{-2 + 3(1+w_m)}}$ &~$\Omega_m(1+w_m)\geq\frac{2}{3}$  & stable \\ \hline
0 & 0 & $\pm\sqrt{1-\Omega_m}$ & $\sqrt{\frac{2(\Omega_C -1)+ 3\Omega_m(1+w_m)}{-2 + 3(1+w_m)}}$ &~$\Omega_m(1+w_m)\geq\frac{2}{3}$  & unstable \\ \hline
0 & $\pm\sqrt{\frac{2(1-\Omega_C) - 3\Omega_m(1+w_m)}{3}}$ & $\sqrt{\frac{1+2\Omega_C + 3\Omega_m w_m}{3}}$ & 0 &~ $\Omega_m(1+w_m)\leq\frac{2}{3}(1-\Omega_C)$ & saddle \\ \hline
1 & 0 & $\pm\sqrt{1-\Omega_m}$ & $-\sqrt{\frac{2(\Omega_C -1)+ 3\Omega_m(1+w_m)}{-2 + 3(1+w_m)}}$ &~$\Omega_m(1+w_m)\geq\frac{2}{3}$  & stable \\ \hline
1 & 0 & $\pm\sqrt{1-\Omega_m}$ & $\sqrt{\frac{2(\Omega_C -1)+ 3\Omega_m(1+w_m)}{-2 + 3(1+w_m)}}$ &~$\Omega_m(1+w_m)\geq\frac{2}{3}$  & unstable \\ \hline
1 & $\pm\sqrt{\frac{2(1-\Omega_C) - \Omega_m(1+w_m)}{6}}$ & $\sqrt{\frac{4+2\Omega_C + 3\Omega_m (w_m-1)}{6}}$ & 0 &~ $\Omega_m(1+w_m)\leq 2(1-\Omega_C)$ & saddle \\ \hline
$\frac{1+w_m\Omega_m}{3(\Omega_m-1)}$ & $\pm\sqrt{\frac{2(1-\Omega_C) - 3\Omega_m(1+w_m)}{2-3\Omega_m(1+w_m)}}$ & $\sqrt{\frac{2(1-\Omega_m)\Omega_C}{2-3\Omega_m(1+w_m)}}$ & 0 &~ $w_m<0,~\Omega_m>\frac{1}{3}$,\newline  $\frac{2}{3}>\Omega_m(1+w_m)$ & unstable \\ \hline
\end{tabularx}
\caption{A table categorising the stability of fixed points found by including a negavive cosmological constant. We assume $0<\Omega_C<1$, disregarding any fixed results that set $\Omega_C=0$ as this reduces back to \cref{tab:power}.}
\end{table}

\begin{figure}
    \centering
        \includegraphics[width=\textwidth]{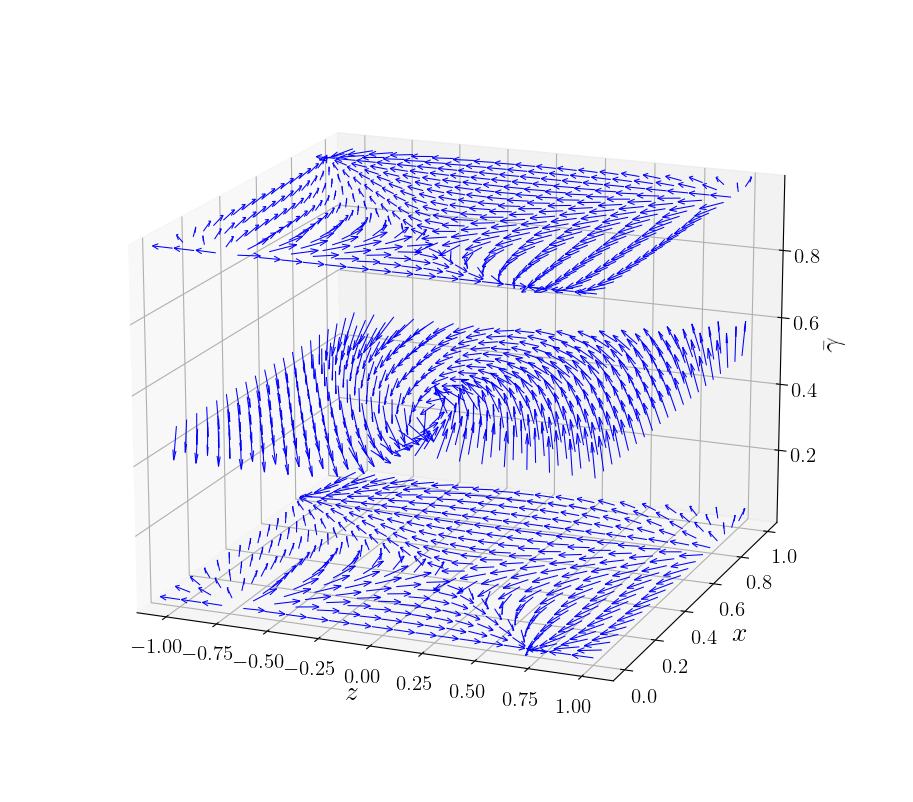}
    \caption{Phase plots for the power-law model ($q=1/2$) with a negative cosmological constant, set $\Omega_C=0.3$. All other parameters are the same used to generate \cref{fig dynamical_sys} }
    \label{fig dyn_sys_lamb}
\end{figure}
As shown in Fig. \ref{fig dyn_sys_lamb}, the introduction of a negative cosmological constant leads to a noteworthy outcome — namely, an expansion in the range of larger '$x$' values that lead to a collapse, as compared to the scenario presented in Fig. \ref{fig dynamical_sys}. This expansion is attributed to the downward shift of the saddle point at $\bar\gamma=0,1$ concerning its '$x$' position, a change documented in \cref{tab:power}. Notably, the fixed points associated with $\bar\gamma=0$ and $\bar\gamma=1$ no longer display asymptotic quasi-de-Sitter solutions at $z=-1$ and $z=1$, as shown in Fig. \ref{fig dyn_sys_lamb} and in \cref{tab:power}. However, this shift of the saddle point towards lower '$x$' values reduces the available parameter space conducive to a successful bounce scenario compared to the scenario discussed in \cref{sec power}.

Our analysis results shows that additional features or advantages emerge from including a negative cosmological constant. But we conclude that including a negative cosmological constant to the DBI model does not exhibit any further enhancement towards a cyclic model.

\section{Conclusion}
\label{sec conc}

DBI models have generated significant interest due to their non-canonical kinetic term introduced by the brane, which results in a unique deceleration mechanism. This mechanism imposes a Lorentz factor that effectively sets a "speed limit" for the DBI scalar field, determined by the brane tension. This deceleration mechanism has led to many attractive inflationary models with unique potentials and non-trivial results. 
More recently, DBI models have been studied in the context of dark energy with interesting and future detectable features. 

The authors of \cite{Copeland:2010jt} conducted an in-depth dynamical analysis of a DBI field regarding dark energy and inflation, analysing both exponential and power-law potentials and brane tensions. We have extended their analysis to include a closed universe, allowing us to explore the stability of additional fixed points, most interestingly around $z=0$. Motivated by string theory phenomenology, we focused our numerical analysis on the power-law scenario, $V \propto \phi^2$ and $f \propto \phi^{-4}$. 

From our analysis, we find the system is unstable in the regime $0<\bar\gamma<1$ and is driven to either the ultra-relativistic case, $\bar\gamma=0$, or the conformal scalar field case, $\bar\gamma=1$. This behaviour is depicted in Fig. \ref{fig dynamical_sys}, where the regime $0<\bar\gamma<1$ drives the system to settle at the fixed points $\bar\gamma=0$ or $\bar\gamma=1$. The fixed point around $z=0$ is a saddle in both the ultra-relativistic and conformal regimes. Although the system still accounts for a bouncing scenario, dependent on initial conditions, it lacks a generic cyclic behaviour. We also find that including a DBI field increases the number of initial conditions that result in a bouncing universe, which reaches a maximum when $\bar\gamma=0$. This result is depicted in Fig. \ref{fig dyn_IC}, where we see that given an ultra-relativistic case, the system has an increased number of initial conditions leading to a bounce compared to a conformally kinetic field.

Motivated by an increase in the initial conditions that allow for a successful bounce, we include an additional degree of freedom to induce a cyclic behaviour. In \cref{sec neg cosmo const}, we investigate the effects of incorporating a negative cosmological constant into the analysis. While introducing a negative cosmological constant expands the range of points leading to collapse, the initial conditions that result in the bounce are constrained by the saddle point. Given a negative cosmological constant, it reduces the $ x$ value of the saddle fixed point. Therefore, in interest to this work, including a negative cosmological constant only reduces the initial conditions that lead to a bounce. Consequently, while a DBI field does lead to an increase of the initial conditions leading to a bounce, a more exotic degree of freedom is required to achieve a generic cyclic behaviour.

\section*{Acknowledgements}
RD is supported by a STFC CDT studentship. CvdB is supported (in part) by the Lancaster–Sheffield Consortium for Fundamental Physics under STFC grant: ST/X000621/1. PKSD thanks the First Rand Bank (SA) for financial support. MC is supported by the National Research Foundation (SA) Scarce-Skills PhD Scholarship, and the University of Cape Town Science Faculty PhD Scholarship.

%%%%%%%%%%%%%%%%%%%%%%%%%%%%%%%%%%%%%%%%%%%%%%%%%%%%%%%%%%%%%%%%


\begin{thebibliography}{9}

\bibitem{Kolb:1990vq} E.~W.~Kolb and M.~S.~Turner, \textit{The Early Universe}, Vol. 69 (1990).

\bibitem{Macias:2017fpu} A.~Mac\'\i{}as and E.~Castellanos, \textit{14th Marcel Grossmann Meeting on Recent Developments in Theoretical and Experimental General Relativity, Astrophysics, and Relativistic Field Theories}, Vol. 1 (2017) pp. 536-547.

\bibitem{Green:1987sp} M.~B.~Green, J.~H.~Schwarz, and E.~Witten, \textit{Superstring Theory. vol. 1: Introduction}, Cambridge Monographs on Mathematical Physics (1988).

\bibitem{Marchesano:2024gul} F.~Marchesano, G.~Shiu, and T.~Weigand, \textit{The Standard Model from String Theory: What Have We Learned?}, 2401.01939.

\bibitem{Liddle:2000cg} A.~R.~Liddle and D.~H.~Lyth, \textit{Cosmological inflation and large scale structure} (2000).

\bibitem{Amendola:2015ksp} L.~Amendola and S.~Tsujikawa, \textit{Dark Energy: Theory and Observations}, Cambridge University Press (2015).

\bibitem{Kachru:2003sx} S.~Kachru, R.~Kallosh, A.~D.~Linde, J.~M.~Maldacena, L.~P.~McAllister, and S.~P.~Trivedi, JCAP {\bf 10}, 013 (2003).

\bibitem{Alishahiha:2004eh} M.~Alishahiha, E.~Silverstein, and D.~Tong, Phys. Rev. D {\bf 70}, 123505 (2004).

\bibitem{Silverstein:2003hf} E.~Silverstein and D.~Tong, Phys. Rev. D {\bf 70}, 103505 (2004).

\bibitem{Armendariz-Picon:1999hyi} C.~Armendariz-Picon, T.~Damour, and V.~F.~Mukhanov, Phys. Lett. B {\bf 458}, 209 (1999).

\bibitem{Garriga:1999vw} J.~Garriga and V.~F.~Mukhanov, Phys. Lett. B {\bf 458}, 219 (1999).

\bibitem{Li:2012vta} S.~Li and A.~R.~Liddle, JCAP {\bf 10}, 011 (2012).

\bibitem{Chimento:2007es} L.~P.~Chimento and R.~Lazkoz, Gen. Rel. Grav. {\bf 40}, 2543 (2008).

\bibitem{Chimento:2010un}  L.~P.~Chimento, R.~Lazkoz, and M.~G.~Richarte, Phys. Rev. D {\bf 83}, 063505 (2011).

\bibitem{Kinney:2007ag} W.~H.~Kinney and K.~Tzirakis, Phys. Rev. D {\bf 77}, 103517 (2008).

\bibitem{Bessada:2009pe} D. Bessada, W. H. Kinney, and K. Tzirakis, JCAP {\bf 09}, 031 (2009).

\bibitem{Kecskemeti:2006cg} S. Kecskemeti, J. Maiden, G. Shiu, and B. Underwood, JHEP {\bf 09}, 076 (2006).

\bibitem{Spalinski:2007dv} M. Spalinski, JCAP {\bf 05}, 017 (2007).

\bibitem{Langlois:2008qf} D. Langlois, S. Renaux-Petel, D. A. Steer, and T. Tanaka, Phys. Rev. D {\bf 78}, 063523 (2008).

\bibitem{Langlois:2008wt} D. Langlois, S. Renaux-Petel, D. A. Steer, and T. Tanaka, Phys. Rev. Lett. {\bf 101}, 061301 (2008).

\bibitem{Chen:2005fe} X. Chen, Phys. Rev. D {\bf 72}, 123518 (2005).

\bibitem{Weller:2011ey} J. M. Weller, C. van de Bruck, and D. F. Mota, JCAP {\bf 06}, 002 (2012).

\bibitem{Huang:2007hh} M.-x. Huang, G. Shiu, and B. Underwood, Phys. Rev. D {\bf 77}, 023511 (2008).

\bibitem{Emery:2013yua} J. Emery, G. Tasinato, and D. Wands, JCAP {\bf 05}, 021 (2013).

\bibitem{Ahn:2009xd} C. Ahn, C. Kim, and E. V. Linder, Phys. Rev. D {\bf 80}, 123016 (2009).

\bibitem{Ahn:2009hu} C. Ahn, C. Kim, and E. V. Linder, Phys. Lett. B {\bf 684}, 181 (2010).

\bibitem{Guo:2008sz} Z.-K. Guo and N. Ohta, JCAP {\bf 04}, 035 (2008).

\bibitem{Chimento:2009nj} L. P. Chimento, R. Lazkoz, and I. Sendra, Gen. Rel. Grav. {\bf 42}, 1189 (2010).

\bibitem{Copeland:2010jt} E. J. Copeland, S. Mizuno, and M. Shaeri, Phys. Rev. D {\bf 81}, 123501 (2010).

\bibitem{Gumjudpai:2009uy} B. Gumjudpai and J. Ward, Phys. Rev. D {\bf 80}, 023528 (2009).

\bibitem{Kaeonikhom:2012xr} C. Kaeonikhom, D. Singleton, S. V. Sushkov, and N. Yongram, Phys. Rev. D {\bf 86}, 124049 (2012).

\bibitem{DiValentino:2019qzk}, Nature Astron. {\bf 4}, 196 (2019).

\bibitem{Yang:2022kho} W. Yang, W. Giar\`e, S. Pan, E. Di Valentino, A. Melchiorri, and J. Silk, Phys. Rev. D {\bf 107}, 063509 (2023).

\bibitem{Handley:2019tkm} W. Handley, Phys. Rev. D {\bf 103}, L041301 (2021).

\bibitem{Calderon:2020hoc}  R. Calder\'on, R. Gannouji, B. L’Huillier, and D. Polarski, Phys. Rev. D {\bf 103}, 023526 (2021).

\bibitem{Visinelli:2019qqu} L. Visinelli, S. Vagnozzi, and U. Danielsson, Symmetry {\bf 11}, 1035 (2019).

\bibitem{Biswas:2009fv} T. Biswas and A. Mazumdar, Phys. Rev. D {\bf 80}, 023519 (2009).

\bibitem{Lucchin:1984yf}  F. Lucchin and S. Matarrese, Phys. Rev. D {\bf 32}, 1316 (1985).

\bibitem{Copeland:1997et} E. J. Copeland, A. R. Liddle, and D. Wands, Phys. Rev. D {\bf 57}, 4686 (1998).

\bibitem{Paliathanasis:2015gga}  A. Paliathanasis, M. Tsamparlis, S. Basilakos, and J. D. Barrow, Phys. Rev. D {\bf 91}, 123535 (2015).

\bibitem{Copeland:2006wr} E. J. Copeland, M. Sami, and S. Tsujikawa, Int. J. Mod. Phys. D {\bf 15}, 1753 (2006).

\end{thebibliography}
\end{document}